\documentclass[a4paper,11pt]{article}
\pdfoutput=1 

\usepackage{jheppub} 

\usepackage[T1]{fontenc} 

\usepackage{graphicx}
\usepackage{xspace}
\usepackage{amsmath}
\usepackage{amsfonts}
\usepackage{amssymb}
\usepackage{enumerate}
\usepackage{epsfig}
\usepackage{lineno}

\newcommand{\VA}[3]{\ifthenelse{\equal{#2}{#3}}
{\ensuremath{#1\pm#2}}{\ensuremath{#1\,^{+#2}_{-#3}}}}

\def\slash#1{\setbox0 = \hbox{$#1$}#1\hskip - \wd0\hbox to\wd0{\hss\sl/\/\hss}}

\newcommand{\ksn}{\ensuremath{\mathrm{K_S}}\xspace}

\newcommand{\kln}{\ensuremath{\mathrm{K_L}}\xspace}

\newcommand{\kn}{\ensuremath{\mathrm{K^0}}\xspace}
\newcommand{\knb}{\ensuremath{\mathrm{\bar{K}^0}}\xspace}

\newcommand{\kpp}{\ensuremath{\mathrm{K_+}}\xspace}
\newcommand{\knn}{\ensuremath{\mathrm{K_-}}\xspace}
\newcommand{\knnmp}{\ensuremath{\mathrm{K_{\mp}}}\xspace}

\newcommand{\kppppm}{\ensuremath{\mathrm{\widetilde K_{\pm}}}\xspace}
\newcommand{\kppp}{\ensuremath{\mathrm{\widetilde K_+}}\xspace}
\newcommand{\knnp}{\ensuremath{\mathrm{\widetilde K_-}}\xspace}
\newcommand{\knt}{\ensuremath{\mathrm{\widetilde K_0}}\xspace}
\newcommand{\knbt}{\ensuremath{\mathrm{\widetilde K_{\bar{0}}}}\xspace}
\newcommand{\kntt}{\ensuremath{\mathrm{ K_0}}\xspace}
\newcommand{\knbtt}{\ensuremath{\mathrm{ K_{\bar{0}}}}\xspace}

\newcommand{\bbn}{\ensuremath{\mathrm{B}}\xspace}
\newcommand{\bbnd}{\ensuremath{\mathrm{B_d}}\xspace}

\newcommand{\kknx}{\ensuremath{\mathrm{K_X}}\xspace}
\newcommand{\kkny}{\ensuremath{\mathrm{K_Y}}\xspace}

\newcommand{\kknxb}{\ensuremath{\mathrm{\bar{K}_X}}\xspace}
\newcommand{\kknyb}{\ensuremath{\mathrm{\bar{K}_Y}}\xspace}
\def\CP       {\ensuremath{\mathrm{CP}}\xspace}
\def\CPT       {\ensuremath{\mathrm{CPT}}\xspace}
\def\C       {\ensuremath{\mathrm{C}}\xspace}
\def\P       {\ensuremath{\mathrm{P}}\xspace}
\def\T       {\ensuremath{\mathrm{T}}\xspace}
\newcommand\trule{\rule{0pt}{2.6ex}}

\newcommand{\jprlBase}       {Phys.\ Rev.\ Lett.\xspace}
\newcommand{\jprBase}        {Phys.\ Rev.\xspace}
\newcommand{\jplBase}        {Phys.\ Lett.\xspace}

\newcommand{\npBase}         {Nucl.\ Phys.\xspace}

\newcommand{\npb}       [1]  {\npBase\ B~{\bf #1}}

\newcommand{\plb}       [1]  {\jplBase\ B~{\bf #1}}

\newcommand{\jprl}      [1]  {\jprlBase\ {\bf #1}}
\newcommand{\pr}        [1]  {\jprBase\ {\bf #1}}
\newcommand{\jprd}      [1]  {\jprBase\ D~{\bf #1}}

\RequirePackage{xspace}

\usepackage{relsize}
\def\babar{\mbox{\slshape B\kern-0.1em{\smaller A}\kern-0.1em
    B\kern-0.1em{\smaller A\kern-0.2em R}}}

\def\CP                {\ensuremath{\mathcal{CP}}\xspace}
\def\CPT               {\ensuremath{\mathcal{CPT}}\xspace} 
\def\C       {\ensuremath{\mathcal{C}}\xspace}
\def\P       {\ensuremath{\mathcal{P}}\xspace}
\def\T       {\ensuremath{\mathcal{T}}\xspace}

\title{Probing CPT in transitions \\
with entangled neutral kaons
}
\author[a]{J. Bernabeu}
\author[b,1]{A. Di Domenico\note{Corresponding author.}}
\author[a,c]{P. Villanueva-Perez}
\affiliation[a]{Department of Theoretical Physics, University of Valencia, and \\
IFIC, Univ. Valencia-CSIC, E-46100 Burjassot, Valencia, Spain}
\affiliation[b]{Department of Physics, Sapienza University of Rome, and \\
INFN Sezione di Roma, P.le A.~Moro, 2, I-00185 Rome, Italy}
\affiliation[c]{Paul Scherrer Institut, Villigen, Switzerland}
\emailAdd{jose.bernabeu@uv.es}
\emailAdd{antonio.didomenico@roma1.infn.it}
\emailAdd{pablo.villanueva@uv.es}

\abstract{
{
In this paper we present a novel \CPT symmetry test 
in the neutral kaon system
based, for the first time, on the direct comparison of the probabilities of a transition and its \CPT reverse.
The required interchange of {\it in} $\leftrightarrow$ {\it out} states for a given process is obtained exploiting the Einstein-Podolsky-Rosen correlations of neutral kaon pairs produced at a $\phi$-factory.
The observable quantities have been constructed by 
selecting the two semileptonic decays for flavour tag,
the $\pi\pi$ and $3\pi^0$ decays for \CP tag and the time orderings
of the decay pairs.
The interpretation in terms of the standard Weisskopf-Wigner approach
to this system, directly connects \CPT violation in these observables to the violating $\Re\delta$
parameter in the mass matrix of $\kn-\knb$, a genuine
\CPT violating effect independent of $\Delta \Gamma$ and not requiring the decay as an essential ingredient.
\\Possible
spurious effects induced by \CP violation in the decay and/or a violation
of the $\Delta S= \Delta Q$ rule have been shown to be 
well under control.
The proposed test is thus fully robust, and might 
shed light on possible new \CPT violating mechanisms, or
further improve the precision of the present experimental limits.
It could be implemented 
at the DA$\Phi$NE facility in Frascati, where the KLOE-2 experiment might reach
a statistical sensitivity of $\mathcal{O}(10^{-3})$ on the newly proposed observable quantities.
}
}

\keywords{CPT symmetry, Discrete Symmetries, Neutral Kaons, $\phi$-factory}
\begin{document}
\maketitle
\flushbottom
\section{Introduction \label{sec:introduction}}
\CPT symmetry, i.e. the symmetry under the combination of charge conjugation (\C), 
parity (\P), and time reversal (\T) transformations, 
at present
appears 
to be the only discrete symmetry of Quantum Mechanics respected in Nature experimentally.
This result has a very solid theoretical foundation in
the well known \CPT theorem
\cite{luders,pauli,bell,jost}
(see also \cite{greenberg1,greenberg2,hollands}
for some more recent developments), 
ensuring exact \CPT 
invariance 
for any quantum field theory formulated on flat space-time
 assuming (1) Lorentz invariance, (2) Locality, and (3) 
Hermiticity.
\par
A violation of the \CPT symmetry
would have 
a dramatic impact
on our present theoretical picture 
and would definitely constitute an unambiguous signal 
of a New Physics framework,
thus strongly motivating
both experimental searches and theoretical studies on this subject.
\par
\CPT invariance has been confirmed by all present experimental tests
(see {\it Tests of conservations laws}
and {\it \CPT invariance tests in neutral kaon decay} reviews in~\cite{ref:pdg2010}),
particularly in the neutral kaon system where 
strong limits have been set to a variety of possible \CPT violation 
effects which might 
arose in a quantum gravity scenario
\cite{ref:KLOE06,ref:KLOE08,ref:KLOE10,ref:KLOE14,cplearabc}.
The best limits on the $\delta$ parameter expressing
\CPT violation in the \kn-\knb mixing matrix,
i.e. in the {\it standard} Weisskopf-Wigner approach~\cite{ww} (see also e.g. appendix A of \cite{kabir}, appendix I of 
\cite{nacht}, or \cite{maiani_cp})
are obtained in the CPLEAR
experiment for $\Re\delta$~\cite{cplearred}, and using the Bell-Steinberger relation for $\Im \delta$ \cite{BS,ref:KLOE_BS}, yielding a stringent limit on the difference of mass terms for \kn and \knb: $\left| m_{\kn}-m_{\knb} \right| < 4 \times 10^{-19} ~\hbox{GeV}$ at 95\% c.l.~\cite{ref:pdg2010}.
\par
The \CPT violating probe has been, however, often limited to a difference of masses (and other intrinsic properties)
for a particle and its anti-particle, i.e. to diagonal mass terms.
In many physical 
phenomena the perturbing effect does not appear at first order in perturbation
theory: it would be sufficient that the perturbation breaks a symmetry of the non-perturbed
states.
{
This vanishing effect at first order for the diagonal elements, 
like e.g. the case of the electric dipole moment for \T violation,
is not present for transitions (non-diagonal elements)
~\cite{colloquium}.
}
%
\par
In this paper we discuss  a new kind of \CPT test for transitions in the neutral kaon system
where the exchange of {\it in} and {\it out} states (and \CP conjugation), required for a {direct and genuine} \CPT test, is performed
exploiting 
the entanglement of the kaon pair 
produced at a $\phi$-factory. This methodology 
has been 
recently proposed for a direct test of the \T symmetry
in the same context \cite{tviol}, 
similarly to the one adopted for the 
performed 
test in the \bbn meson system 
at \bbn-factories \cite{ref:bernabeuPLB,ref:bernabeuNPB,ref:Bmethod,ref:babarTviol,colloquium}.
%
The decay is not an essential ingredient for a non-vanishing effect 
and it is only used for
filtering the appropriate initial and final states of the neutral kaon transition~\cite{ref:Wolfenstein}.
Explicitly, in the
standard  Weisskopf-Wigner approach to this system,
our \CPT-violating effects can be
connected to the $\Re \delta$ parameter, 
a genuine \CPT-violating effect independent of $\Delta \Gamma=\Gamma_S-\Gamma_L$,
with $\Gamma_S$ and $\Gamma_L$ the widths of the physical states (see Section \ref{section2}).
\par
%
\par
Using this method it would be possible for the first time to 
directly test the \CPT symmetry in transition processes between meson states, rather than comparing 
masses, lifetimes, or other intrinsic properties of particle and anti-particle states.
Any \CPT test requires a very clear theoretical formulation and very clean experimental 
conditions in order to avoid possible 
fake effects
mimicking a \CPT violation signal, e.g. possible and uncontrolled direct \CP violation contributions
%
%
that could
alter the precise identification of the filtered meson state. One has to be aware that
a significant result in a \CPT test will involve small numbers, so that  a precise control
of the experimental conditions is mandatory. 
%
%
The test proposed in this paper exploits 
the peculiarities
of the neutral kaon system and would allow a very
clean and direct test of the \CPT symmetry.
This is in contrast with the problems encountered in
 the neutral \bbn meson system,
where one has very severe limitations to a possible clean test of \CPT due to the use of the
decay  
$\bbn \rightarrow J/\Psi \kln$, with the current 
identification of $\kln$ which is not an exact \CP
eigenstate.
%
%
%
\par
The KLOE-2 experiment~\cite{kloe2epjc} at DA$\Phi$NE, the Frascati $\phi$-factory, 
might perform for the first time this kind of test 
reaching a sensitivity of $\mathcal{O}(10^{-3})$ in the experimental asymmetry, 
a high sensitivity which - we emphasize - 
is  a direct test by comparing a given transition with its \CPT-transformed transition,
and not a result of fitting a parameter in a given model. If interpreted in terms of the 
sensitivity to the usual \CPT-violating $\delta$ parameter  for the \kn-\knb system,
it would also allow to further improve the present limits on $\delta$.
\par
In Section \ref{section2} we precisely define
the {\it in} and {\it out} kaon
states involved in the time evolution of the system.
In Section \ref{section3} we construct the \CPT-violating observables using appropriate ratios of transition probabilities.
{
In Section \ref{sec:orthogonality} we discuss the impact of the approximations and the orthogonality problem for the states filtered by the decay channel, and 
evaluate the sensitivity
of the proposed test.
}
Finally in Section \ref{conclusions} we present our conclusions.

\section{Definition of states}
\label{section2}
\par
In order to formulate a possible \CPT symmetry test with neutral kaons,
it is necessary to precisely define the different states involved as possible 
{\it in} and {\it out} states in the time evolution. 
First, let us consider the short- and long-lived 
physical states $|\ksn\rangle$ and $|\kln\rangle$, 
i.e. the states with definite masses $m_{S,L}$ and lifetimes $\tau_{S,L}$
which evolve as a function of the kaon proper time $t$ as pure exponentials 
\begin{linenomath*}
\begin{eqnarray}
|\ksn(t)\rangle&=&e^{-i\lambda_{S}t}|\ksn\rangle \nonumber\\
|\kln(t)\rangle&=&e^{-i\lambda_{L}t}|\kln\rangle ~.
\end{eqnarray}
\end{linenomath*}

with $\lambda_{S,L}=m_{S,L}-i\Gamma_{S,L}/2$, and $\Gamma_{S,L}=(\tau_{S,L})^{-1}$.
They are usually expressed in terms of the flavor 
eigenstates $|\kn\rangle$, $|\knb\rangle$ as:
\begin{linenomath*}
\begin{eqnarray}
|\ksn\rangle &=& \frac{1}{\sqrt{2\left(1+|\epsilon_S|^2\right)}}
\left[ (1+\epsilon_S) |\kn \rangle
+(1-\epsilon_S) |\knb \rangle
\right] \\
|\kln\rangle &=& \frac{1}{\sqrt{2\left(1+|\epsilon_L|^2\right)}}
\left[ (1+\epsilon_L) |\kn \rangle
-(1-\epsilon_L) |\knb \rangle
\right]
~,
\end{eqnarray}
\end{linenomath*}

with $\epsilon_S$
and $\epsilon_L$
two small complex parameters describing the \CP impurity in the physical states.
One can equivalently define
${\epsilon} \equiv (\epsilon_S+\epsilon_L)/2$, and
$ \delta \equiv (\epsilon_S-\epsilon_L)/2$;
adopting a suitable phase convention 
(e.g. the Wu-Yang phase convention \cite{ref:wuyang}) 
$\epsilon\neq0$ implies \T violation,
$\delta\neq0$ implies \CPT violation,
while $\delta\neq0$ or
$\epsilon\neq0$ implies \CP violation.
\par
Let us also
consider the states $|\kpp\rangle$, $|\knn\rangle$ defined as follows:
$|\kpp\rangle$ is the state filtered by the decay into $\pi\pi$ 
($\pi^+\pi^+$ or $\pi^0\pi^0$), a
pure $\CP=+1$ state; analogously $|\knn\rangle$ is the state filtered by the decay into 
$3\pi^0$, a
pure $\CP=-1$ state. 
Their orthogonal states correspond to the states which cannot decay into 
$\pi\pi$ or $3\pi^0$, defined, respectively, as
\begin{linenomath*}
\begin{eqnarray}
|\knnp\rangle &\equiv & {\rm\widetilde N_-} \left[| \kln\rangle 
- \eta_{\pi\pi} |\ksn \rangle \right] 
\\
|\kppp\rangle &\equiv & {\rm\widetilde N_+} \left[| \ksn\rangle 
- \eta_{3\pi^0} |\kln \rangle \right] 
\end{eqnarray}
\end{linenomath*}
%
with 
\begin{eqnarray}
\eta_{\pi\pi}&=&\frac{\langle \pi\pi |T |\kln\rangle}{\langle \pi\pi |T |\ksn\rangle}
\\
\eta_{3\pi^0}&=&\frac{\langle 3\pi^0 |T |\ksn\rangle}{\langle 3\pi^0 |T |\kln\rangle}~, 
\end{eqnarray}
and
${\rm\widetilde N_{\pm}}$ two suitable normalization factors.
With these definitions of states, $|\kpp\rangle$ and $|\knn\rangle$ can be explicitly
constructed imposing the conditions
$\langle\kppppm|\knnmp\rangle=0$:
\begin{linenomath*}
\begin{equation}
|\kpp\rangle = {\rm N_+} \left[| \ksn\rangle 
+ \alpha |\kln \rangle \right]
\end{equation}
\end{linenomath*}
\begin{linenomath*}
\begin{equation}
|\knn\rangle = {\rm N_-} \left[| \kln\rangle 
+ \beta |\ksn \rangle \right]
\end{equation}
\end{linenomath*}
where 
\begin{linenomath*}
\begin{equation}
\alpha=\frac{\eta_{\pi\pi}^{\star}-\langle \kln|\ksn\rangle}
{1-\eta_{\pi\pi}^{\star}\langle \ksn|\kln\rangle}~,
\end{equation}
\end{linenomath*}
\begin{linenomath*}
\begin{equation}
\beta=\frac{ \eta_{3\pi^0}^{\star}-\langle \ksn|\kln\rangle}
{1-\eta_{3\pi^0}^{\star} \langle \kln|\ksn\rangle} ~, 
\end{equation}
\end{linenomath*}
and $\rm N_{\pm}$ are two normalization factors.
\par 
Here we have kept separate definitions of 
the {\it filtered} states $|\kpp\rangle$ and $|\knn\rangle$,
which are observed
through their decay,
from the {\it tagged} states $|\kppp\rangle$ and $|\knnp\rangle$,
which are prepared exploiting 
the entanglement of the
kaon pairs, 
as we will discuss 
in the next section.
$|\kpp\rangle$ and $|\knn\rangle$ are defined as the filtered states when observing 
definite $\CP=\pm 1$ decay products. Even though the decay products are orthogonal, the filtered 
$|\kpp\rangle$ and $|\knn\rangle$ states can still be nonorthoghonal.
In the following we will assume
\begin{linenomath*}
\begin{eqnarray}
|\kpp\rangle&\equiv&|\kppp\rangle \nonumber \\
|\knn\rangle&\equiv&|\knnp\rangle~,
\label{eq:equiv}
\end{eqnarray}
\end{linenomath*}
which corresponds to impose
the condition of orthogonality $\langle\knn|\kpp\rangle=0$,
implying
that $\beta = -\eta_{\pi\pi}$ and $\alpha=-\eta_{3\pi^0}$, and
a precise relationship between the two amplitude ratios $\eta_{\pi\pi}$ 
and $\eta_{3\pi^0}$:
\begin{linenomath*}
\begin{eqnarray}
\eta_{\pi\pi}+\eta_{3\pi^0}^{\star} - \eta_{\pi\pi} \eta_{3\pi^0}^{\star} \langle \kln|\ksn\rangle
&=& \langle \ksn|\kln\rangle 
\nonumber\\
&=&
\frac
{\epsilon_L + \epsilon_S^{\star}}
{\sqrt{(1+|\epsilon_L|^2)(1+|\epsilon_S|^2) }}
\label{eq:etass}
~,
\end{eqnarray}
\end{linenomath*}
Neglecting terms of $\mathcal{O}(\epsilon^3)$
(with $\epsilon = 
\mathcal{O}(10^{-3})$), therefore
with a high degree of accuracy,
$\mathcal{O}(10^{-9})$,
this translates into the following relation:
\begin{linenomath*}
\begin{eqnarray}
\eta_{\pi\pi}+\eta_{3\pi^0}^{\star}=\epsilon_L+\epsilon_S^{\star}~.
\label{eq:etass2}
\end{eqnarray}
\end{linenomath*}
This 
clearly indicates
that direct \CP and \CPT violation have to be neglected when imposing assumption
(\ref{eq:equiv}).
 In fact, for instance, eq.(\ref{eq:etass2}) 
cannot be simultaneously satisfied for $\pi^+\pi^-$ and $\pi^0\pi^0$
decays, being $(\eta_{\pi^+\pi^-}-\eta_{\pi^0\pi^0})=3\epsilon^{\prime}$, 
with $\epsilon^{\prime}= \mathcal{O}(10^{-6})$ the direct \CP violation parameter \cite{ref:pdg2010}.
Similar subtle points were previously
discussed in the literature for the \T-asymmetry measurement in the
flavour-\CP eigenstates of $J/\Psi \kn$ 
decay channels of \bbnd's~\cite{ref:nir}, as
well as for any pair of decay channels~\cite{ref:botella}.
{
\\ 
More in general, while possible direct \CPT violation contributions might be still cast into the definition of the observable quantities for the 
\CPT test that will be presented in the next Section,
direct \CP violation may appear as a contaminating fake effect which is necessary to keep well under control.
}
%
\par
Finally the validity of the $\Delta S=\Delta Q$ rule will be assumed in the following, so that the two flavor orthogonal eigenstates $|\kn\rangle$ and $|\knb\rangle$ are identified by the charge of the lepton in semileptonic decays.
When the decay into $\pi^-\ell^+\nu$  is observed, it cannot come from $|\knb\rangle$ so that 
the state $|\kn\rangle$ is filtered, and vice-versa for the decay into $\pi^+\ell^-\bar{\nu}$.
\par
The relevance of these assumptions  will be discussed in Section
\ref{sec:orthogonality}, where it will be shown that 
they can be safely released for our purposes, 
without
affecting the cleanliness of the test.
%
%
%
\section{\CPT symmetry test at a $\phi$-factory}
\label{section3}
Similarly to the \T symmetry test proposed at a $\phi$-factory (or B-factory)
 \cite{tviol,ref:bernabeuPLB,ref:bernabeuNPB,ref:Bmethod}, the implementation of the \CPT test proposed here
 exploits the 
 Einstein-Podolsky-Rosen (EPR) 
\cite{ref:EPR}
 entanglement of the neutral meson pair produced in $\phi\rightarrow \kn\knb$ decays.
In fact in this case
the initial state 
of the pair is totally antisymmetric\footnote{
It is worth noting that a possible \CPT violation effect
in the entangled state,
the so-called $\omega$-effect~\cite{mavro1,mavro2},
might induce a breakdown of the 
antisymmetry 
of the initial state. 
The possible impact of this effect, which
is 
strongly
bounded
experimentally 
\cite{ref:KLOE06,ref:KLOE08,ref:KLOE10},
is not considered in this paper, for simplicity.
}
and
can be written in terms of any pair of orthogonal states, e.g. \kn and \knb, or  \kpp and \knn,  as:
\begin{linenomath*}
\begin{eqnarray}
  |i \rangle   =  \frac{1}{\sqrt{2}} \{ |\kn \rangle |\knb \rangle - 
 |\knb \rangle |\kn \rangle
\} 
   =  \frac{1}{\sqrt{2}} \{ |\kpp \rangle |\knn \rangle - 
 |\knn \rangle |\kpp \rangle
\label{eq:state1}
\}~.
\end{eqnarray}  
\end{linenomath*}
%
Thus, exploiting the perfect anticorrelation of the state implied 
by eq.~(\ref{eq:state1}), which remains unaltered -- even in the presence of \kn-\knb mixing -- until one of the two kaons decays,
 it is possible to have a 
\textquotedblleft flavor-tag\textquotedblright~or a 
 \textquotedblleft \CP-tag\textquotedblright,
i.e.~to prepare a definite
\kn or \knb state,
 or a definite 
 \kpp or \knn state
of the still alive kaon by observing a specific flavor decay\footnote{In the following the semileptonic decays $\pi^+\ell^-\nu$
and  $\pi^-\ell^+\bar{\nu}$ will be 
denoted 
for brevity
as $\ell^-$ and $\ell^+$, respectively.} 
($\ell^-$ or $\ell^+$)
or \CP decay ($3\pi^0$ or $\pi\pi$ ) of the other (and first decaying) kaon in the pair.
For instance, the transition $\kn \to \kpp$ and its 
associated probability $P\left[\kn(0)\to\kpp(\Delta t)\right]$
corresponds to the
observation of a $\ell^{-}$ decay at a proper time $t_1$ of the opposite \knb 
and a $\pi\pi$ decay at a later proper time $t_2=t_1+\Delta t$, with $\Delta t >0$. 
\\In other words,
the $\ell^{-}$ decay of a kaon on one side prepares (tags), in the quantum mechanical sense, the opposite (if undecayed) kaon
in the state $|\kn \rangle$ at a starting time $t=0$. The $|\kn\rangle$ state freely evolves in time until its $\pi\pi$ decay filters 
it in the state $|\kpp \rangle$ at a time $t=\Delta t$.
%
%
%
\par
In this way 
one can experimentally access all the four reference transitions listed in Table~\ref{tab:processes},
and their \T, \CP and \CPT conjugated transitions.
It can be easily checked that the three conjugated transitions
correspond to different categories of events; therefore the comparisons between reference vs conjugated transitions
correspond to independent \T, \CP and \CPT experimental tests.
None of these transformed transitions for the three symmetries 
are simply
an exchange of the time ordering
of the two decay channels.
\begin{table}[h]
  \begin{center}
    \begin{tabular}{c|c|c|c}		
      \hline
      Reference & \T-conjug.   &  \CP-conjug. & \CPT-conjug. \\ \hline
      \trule $\kn \to \kpp$   & $\kpp \to \kn$    & $\knb \to \kpp$ & $\kpp \to \knb$ \\
      \trule $\kn \to \knn$ &  $\knn \to \kn$ &  $\knb \to \knn$ &  $\knn \to \knb$ \\
      \trule $\knb \to \kpp$   & $\kpp \to \knb$    &  $\kn \to \kpp$ & $\kpp \to \kn$ \\
      \trule $\knb \to \knn$  & $\knn \to \knb$   & $\kn \to \knn$  & $\knn \to \kn$   \\ \hline
   \end{tabular}
    \caption{Scheme of possible reference transitions and their associated
 \T, \CP or \CPT conjugated processes accessible at a $\phi$-factory.
      \label{tab:processes}}
  \end{center}	
\end{table}
%
\par
For the \CPT symmetry test one can define 
the following ratios of probabilities:
\begin{linenomath*}
\begin{eqnarray}
R_{1,\CPT}(\Delta t) &=& \frac{P\left[\kpp(0)\to\knb(\Delta t)\right]}{P\left[\kn(0)\to\kpp(\Delta t)\right]} \nonumber \\
R_{2,\CPT}(\Delta t) &=& \frac{P\left[\kn(0)\to\knn(\Delta t)\right]}{P\left[\knn(0)\to\knb(\Delta t)\right]} \nonumber\\
R_{3,\CPT}(\Delta t) &=& \frac{P\left[\kpp(0)\to\kn(\Delta t)\right]}{P\left[\knb(0)\to\kpp(\Delta t)\right]} \nonumber\\
R_{4,\CPT}(\Delta t) &=& \frac{P\left[\knb(0)\to\knn(\Delta t)\right]}{P\left[\knn(0)\to\kn(\Delta t)\right]}~.
\label{eq:ratios}
\end{eqnarray}
\end{linenomath*}
The measurement of any deviation from the prediction $R_{i,\CPT}(\Delta t)=1$
imposed by \CPT invariance
is a signal of \CPT violation. 
\\
It is worth noting that for $\Delta t=0$:
\begin{linenomath*}
\begin{eqnarray}
R_{1,\CPT}(0)=R_{2,\CPT}(0)=R_{3,\CPT}(0)=R_{4,\CPT}(0)=1~,
\label{eq:one}
\end{eqnarray}
\end{linenomath*}
i.e. the \CPT-violating effect is built in the time 
evolution of the system, and it is absent at $\Delta t=0$, within our approximations.
Possible deviations from limits (\ref{eq:one}) will be discussed in the next Section.
\\
For $\Delta t\gg \tau_S$, assuming the presence of \CPT violation only in the mass matrix
($\delta \neq 0)$ and nothing else, 
one gets:
\begin{linenomath*}
\begin{eqnarray}
R_{2,\CPT}(\Delta t \gg \tau_S) &\simeq& 1-4 \Re \delta
\label{eq:tendtoconst1}
\\
R_{4,\CPT}(\Delta t \gg \tau_S) &\simeq& 1+4\Re \delta~,
\label{eq:tendtoconst2}
\end{eqnarray}
\end{linenomath*}
%
i.e. the \CPT-violating effect built in the time evolution reaches a ``plateau'' regime and dominates
 in this limit. It is a {\it genuine} effect because $\Re \delta$ does
not depend on $\Delta \Gamma$ as an essential ingredient \cite{colloquium,ref:Wolfenstein,ref:Wolfenstein2}. 
This constitutes an important difference with respect to the \T symmetry test case \cite{tviol}, where
the effect due to the $\Re \epsilon$ parameter 
vanishes for $\Delta \Gamma \rightarrow 0$.
%
\par
At a $\phi$-factory the observable quantity
is the
double differential decay rate 
$I(f_1,t_1;f_2,t_2) $
of the state $|i\rangle$
into decay products
$f_1$ 
and $f_2$ 
at proper times $t_1$ and $t_2$, respectively \cite{ref:HandbookAD}.
\par
After integration on $t_1$ at fixed time difference $\Delta t=t_2-t_1\geq 0 $,
the resulting decay intensity
$I(f_1, f_2 ; \Delta t) $
can be rewritten 
in a 
suitable form, putting in evidence the probabilities we are aiming for.
In fact, for two generic pairs of orthogonal states $\{\kknx,\kknxb\}$ and $\{\kkny,\kknyb\}$ 
uniquely identified by the decay products $\{f_X,f_{\bar{X}}\}$ and $\{f_Y,f_{\bar{Y}}\}$ 
(e.g. the $\{\kn,\knb\}$ and $\{\kpp,\knn\}$ pairs identified by the decay products
$\{\ell^+,\ell^-\}$ and $\{\pi\pi,3\pi^0\}$, respectively, under the assumption of eqs.(\ref{eq:equiv}) and the $\Delta S=\Delta Q$ rule),
%
the decay intensity can be expressed as:
%
\begin{linenomath*}
\begin{eqnarray}
I(f_{\bar{X}},f_{Y};\Delta t)&=& \int^{\infty}_0 I(f_{\bar{X}}, t_1;f_{Y};t_2) d t_1
\nonumber \\
&=&C(f_{\bar{X}},f_Y)\times P\left[\kknx(0) \to \kkny(\Delta t) \right]~,
\end{eqnarray}
\end{linenomath*}
where the coefficient $C(f_{\bar{X}},f_Y)$, depending only on the final states $f_{\bar{X}}$ and $f_Y$, is 
given by:
\begin{linenomath*}
\begin{eqnarray}
C(f_{\bar{X}},f_Y)&=&\frac{1}{2(\Gamma_S+\Gamma_L)} \left|  
\langle f_{\bar{X}} | T |  \kknxb  \rangle
\langle f_Y | T |\kkny  \rangle
\right|^2 
\label{eq:coeffi}
\end{eqnarray}
\end{linenomath*}
and $P\left[\kknx(0) \to \kkny(\Delta t) \right]$  is the generic $\kknx \to \kkny$ transition probability which contains the 
$\Delta t$ time dependence only.
\\
It's worth noting that a similar expression can be easily formulated also for the case $\Delta t <0$:
\begin{linenomath*}
\begin{eqnarray}
I(f_{\bar{X}},f_{Y};\Delta t)
=C(f_{\bar{X}},f_Y)\times P\left[\kknyb(0) \to \kknxb(|\Delta t|) \right]~.
\end{eqnarray}
\end{linenomath*}
\par
Therefore, at a $\phi$-factory
one can define the observable ratios:
\begin{linenomath*}
\begin{eqnarray}
\label{ratio2}
R_{2,\CPT}^{\rm{exp}}(\Delta t) &\equiv&
\frac{  I(\ell^-,3\pi^0;\Delta t)}
{ I(\pi\pi,\ell^-;\Delta t)}
\end{eqnarray} 
\end{linenomath*}
\begin{linenomath*}
\begin{eqnarray}
\label{ratio4}
R_{4,\CPT}^{\rm{exp}}(\Delta t) &\equiv&
\frac{  I(\ell^+,3\pi^0;\Delta t)}
{ I(\pi\pi,\ell^+;\Delta t)}   ~,
\end{eqnarray} 
\end{linenomath*}
which are related to the $R_{i,\CPT}(\Delta t)$ ratios defined in eqs (\ref{eq:ratios}) as follows, for $\Delta t  \geq 0$:
\begin{linenomath*}
\begin{eqnarray}
\label{eq:intensity2}
R_{2,\CPT}^{\rm{exp}}(\Delta t) 
&=&{R_{2,\CPT}(\Delta t)}\times{D_{\CPT}} 
\nonumber\\
R_{4,\CPT}^{\rm{exp}}(\Delta t) 
&=&{R_{4,\CPT}(\Delta t)}\times{D_{\CPT}}
\end{eqnarray}  
\end{linenomath*}
whereas
for $\Delta t <0$ one has:
\begin{linenomath*}
\begin{eqnarray}
\label{eq:intensity2}
R_{2,\CPT}^{\rm{exp}}(\Delta t) 
&=&{R_{1,\CPT}(|\Delta t|)}\times{D_{\CPT}}
\nonumber\\
R_{4,\CPT}^{\rm{exp}}(\Delta t) 
&=&{R_{3,\CPT}(|\Delta t|)}\times{D_{\CPT}}~,
\end{eqnarray}  
\end{linenomath*}
with $D_{\CPT}$ the ratio of coefficients:
\begin{linenomath*}
\begin{eqnarray}  
 D_{\CPT}= \frac{  C(\ell^-,3\pi^0;\Delta t)}
{ C(\pi\pi,\ell^-;\Delta t)}  = \frac{  C(\ell^+,3\pi^0;\Delta t)}
{ C(\pi\pi,\ell^+;\Delta t)} 
= \frac{\left| 
\langle 3\pi^0|T| \knn \rangle\right|^2}{ \left|\langle \pi\pi|T| \kpp \rangle\right|^2}
\end{eqnarray}  
\end{linenomath*}
that can be expressed, with a high degree of accuracy, at least $\mathcal{O}(10^{-7})$, as:
\begin{linenomath*}
\begin{eqnarray}  
D_{\CPT}=
\frac{ {\rm BR} \left( \kln\rightarrow 3\pi^0\right) }{ {\rm BR}\left( \ksn\rightarrow \pi\pi\right)}
\frac{\Gamma_L}{\Gamma_S}~.
\end{eqnarray}   
\end{linenomath*}
\par
The value of $D_{\CPT}$ can be
therefore evaluated 
from branching ratios and lifetimes, 
but it is also directly measurable 
from the observable ratios (\ref{ratio2}) and  (\ref{ratio4}),
as it will be discussed in detail in the next Section.
\par
The explicit expressions of ratios (\ref{ratio2}) and  (\ref{ratio4}) (neglecting higher order terms in small parameters
and for not too large negative $\Delta t$) are:
\begin{linenomath*}
\begin{eqnarray}
R_{2,\CPT}^{\rm{exp}} ( \Delta t ) &=&
\frac{P[\kn(0)\to\knn(\Delta t)]}{P[\knn(0)\to\knb(\Delta t)]} \times D_{\CPT} 
\nonumber\\
&\simeq& |1-2\delta|^2 \left| 1+ 2\delta 
e^{-i(\lambda_S-\lambda_L)\Delta t} \right|^2
\times D_{\CPT}~,
\label{ratio2expl}
\end{eqnarray}
\end{linenomath*}
\begin{linenomath*}
\begin{eqnarray}
R_{4,\CPT}^{\rm{exp}} ( \Delta t ) &=&
\frac{P[\knb(0)\to\knn(\Delta t)]}{P[\knn(0)\to\kn(\Delta t)]} \times D_{\CPT}
\nonumber\\
&\simeq& |1+2\delta|^2 \left| 1- 
2\delta 
e^{-i(\lambda_S-\lambda_L)\Delta t} \right|^2
\times D_{\CPT}~.
\label{ratio4expl}
\end{eqnarray}
\end{linenomath*}
\par
\begin{figure}[htbp] 
   \centering
   \includegraphics[width=5in]{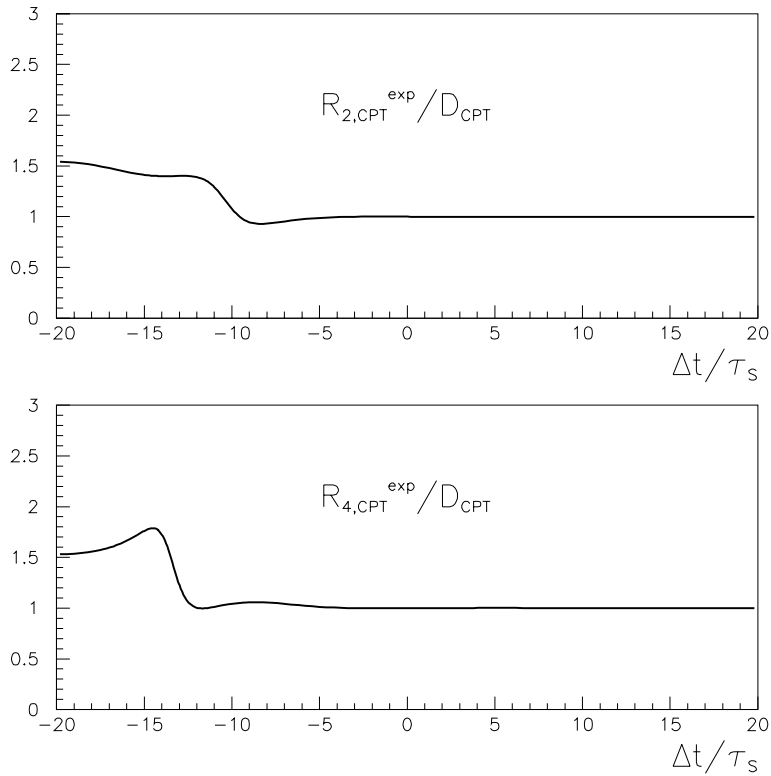} 
   \caption{The ratios $R_{2,\CPT}^{\rm{exp}}(\Delta t)$ and $R_{4,\CPT}^{\rm{exp}}(\Delta t)$
as a function of $\Delta t$. For visualization purposes 
 the \CPT violating parameters have been fixed to the values $\Re \delta=3.3 \times 10^{-4}$ and $\Im \delta = 1.6 \times 10^{-5}$~.}
   \label{fig1}
\end{figure}
\begin{figure}[htbp] 
   \centering
   \includegraphics[width=5in]{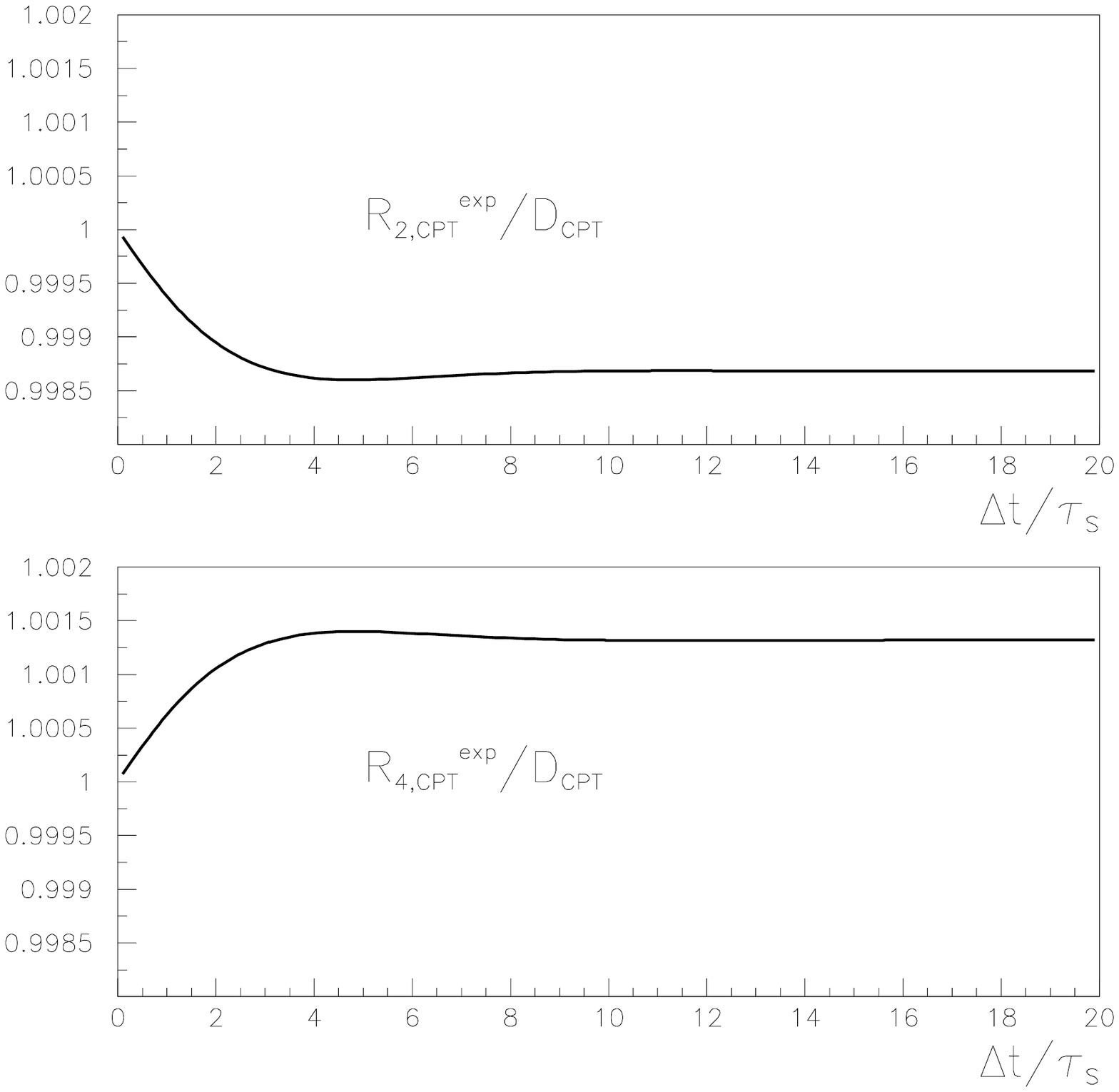} 
   \caption{A zoom of the plots shown in Fig.\ref{fig1} in the region $0\leq \Delta t \leq20 \tau_S$.
}
   \label{fig2}
\end{figure}

{
The 
 expected behavior of the observables $R_{2,\CPT}^{\rm{exp}}(\Delta t)$ and $R_{4,\CPT}^{\rm{exp}}(\Delta t)$ 
as a function of $\Delta t$, and without the approximations of eqs.(\ref{ratio2expl}) and (\ref{ratio4expl}), is shown in Fig.\ref{fig1}, where -- for visualization purposes -- 
the probabilities involved have been evaluated fixing the \CPT violating parameters $\Re \delta$ and $\Im \delta$ to a value different from zero, and 
equal to their present uncertainties \cite{ref:pdg2010}, i.e. $\Re \delta=3.3 \times 10^{-4}$ and $\Im \delta = 1.6 \times 10^{-5}$~.
In Fig.\ref{fig2} a zoom 
of the $\Delta t >0$ region, where the 
\textquotedblleft plateau\textquotedblright~
regimes (\ref{eq:tendtoconst1}) and (\ref{eq:tendtoconst2}) dominate,
is shown.
Experimentally, this is the most interesting and statistically most populated region, where the best
sensitivity to \CPT violation effects can be reached by the KLOE-2 experiment
(see Section  \ref{sec:orthogonality}).
%
}

\par
We emphasise that these observables are genuine \CPT violating effects by comparing experimentally the probability for a given transition and its \CPT reverse, independent of any theoretical scenario generating this effect.
When they are interpreted in a model for \CPT violation in the mass matrix (i.e. with $\delta \neq 0$) and nothing else, these observables can be compared with the result expected for the survival probabilities (diagonal processes) like the one that  has been measured by the CPLEAR experiment~\cite{cplearred}.
In this case, the CPLEAR asymmetry can be easily translated into our formalism as an observable ratio
of probabilities (in this case $D_{\CPT}=1$):
\begin{linenomath*}
\begin{eqnarray}
\label{cplearratio}
\frac{  I(\ell^-,\ell^+;\Delta t)}
{ I(\ell^+,\ell^-;\Delta t)} &=&
\frac{P[\kn(0)\to\kn(\Delta t)]}{P[\knb(0)\to\knb(\Delta t)]} 
\nonumber\\
&\simeq& |1-4\delta|^2 \left| 1+ \frac{8 \delta}{1+e^{+i(\lambda_S-\lambda_L)\Delta t}} \right|^2 ~.
\end{eqnarray}
\end{linenomath*}

The comparison of equations (\ref{ratio2expl}) and (\ref{ratio4expl}) with equation (\ref{cplearratio}) shows that even within the {\it same} model,
the difference in the dependence on the \CPT violation parameter $\delta$ is apparent.
In the limit $\Delta \Gamma \rightarrow 0$  the ratio (\ref{cplearratio}) 
tends to unity for all times,
whereas ratios (\ref{ratio2expl}) and (\ref{ratio4expl}) 
are different from unity
through $\Re \delta$, which is
independent of $\Delta \Gamma$.
Moreover, just as an illustration of the different sensitivity of these observables to \CPT violation,
in the hypothesis of 
\CPT-violating effects introduced via a deviation from conventional quantum mechanics, 
believed to reflect the loss of quantum coherence expected in some approaches to quantum gravity
\cite{ellis1,ellis2},
the ratio (\ref{cplearratio}) is insensitive to these effects
(up to second order in the \CPT violation parameters
of the model and for all times~\cite{ellis2}), 
while it can be shown that ratios (\ref{ratio2expl}) and (\ref{ratio4expl}) 
are sensitive to them
at first order in some of the parameters.
%
%
\section{Impact of the approximations on the test. Results.}
\label{sec:orthogonality}

\par
{
In order to study the impact 
of the approximations involved in the proposed \CPT test,
namely 
negligible direct \CP and \CPT violation contributions in the $\pi\pi$ and $3\pi^0$ channels,
and the validity of the $\Delta S=\Delta Q$ rule,
they are treated separately.
\par
First, the effect of possible direct \CP and \CPT violation contributions
is evaluated
on the observable ratios $R_{i,\CPT}^{\rm{exp}}(\Delta t)$,
while still assuming the $\Delta S=\Delta Q$ rule. 
To this aim the following parametrisation is introduced:
\begin{linenomath*}
\begin{eqnarray}
\eta_{\pi\pi}&=&\epsilon_L+\epsilon^{\prime}_{\pi\pi} \nonumber\\
\eta_{3\pi^0}&=&\epsilon_S+\epsilon^{\prime}_{3\pi^0}~,
\end{eqnarray}
\end{linenomath*}
where $\epsilon^{\prime}_{\pi\pi}$ and $\epsilon^{\prime}_{3\pi^0}$ represent the generic contributions of direct \CP and/or \CPT violation
in the $\pi\pi$ and $3\pi^0$ channels, respectively.
In this more general case, the orthogonality condition 
eqs(\ref{eq:equiv})
is no more satisfied,
and the true orthogonal pair to be considered in writing the initial state (\ref{eq:state1}) is 
$\{\kpp,\knnp\}$ (or $\{\kppp,\knn\}$) instead of $\{\kpp,\knn\}$.
}
%
%
The effect of $\epsilon^{\prime}_{\pi\pi}$ and $\epsilon^{\prime}_{3\pi^0}$ 
can be easily singled out in the explicit expressions of the observable ratios (neglecting higher order terms in small parameters
and for not too large negative $\Delta t$):
\begin{linenomath*}
\begin{eqnarray}
R_{2,\CPT}^{\rm{exp}} ( \Delta t ) &=&
\frac{P[\kn(0)\to\knn(\Delta t)]}{P[\knnp(0)\to\knb(\Delta t)]} \times D_{\CPT} 
\nonumber\\
&=& \frac{\left|e^{-i\lambda_S\Delta t}\left(\frac{1-\epsilon_L}{\sqrt{2}}\right)
\eta_{3\pi^0}
+e^{-i\lambda_L\Delta t}
\left(\frac{1-\epsilon_S}{\sqrt{2}}\right)\right|^2}
{\left|e^{-i\lambda_S\Delta t}\left(\frac{1-\epsilon_S}{\sqrt{2}}\right)
{\eta_{\pi\pi}}
+e^{-i\lambda_L\Delta t}
\left(\frac{1-\epsilon_L}{\sqrt{2}}\right)\right|^2} 
\times D_{\CPT}~,
\nonumber\\
&\simeq& |1-2\delta|^2 \left| 1+ 
\left(
\eta_{3\pi^0}-{\eta_{\pi\pi}}
%
\right)
e^{-i(\lambda_S-\lambda_L)\Delta t} \right|^2
\times D_{\CPT}~,
\nonumber\\
&=& |1-2\delta|^2 \left| 1+ 
\left(2\delta 
+\epsilon^{\prime}_{3\pi^0} -\epsilon^{\prime}_{\pi\pi}  \right)
e^{-i(\lambda_S-\lambda_L)\Delta t} \right|^2
\times D_{\CPT}~,
\label{ratio2prime}
\end{eqnarray}
\end{linenomath*}
\begin{linenomath*}
\begin{eqnarray}
R_{4,\CPT}^{\rm{exp}} ( \Delta t ) &=&
\frac{P[\knb(0)\to\knn(\Delta t)]}{P[\knnp(0)\to\kn(\Delta t)]} \times D_{\CPT}
\nonumber\\
&=&
\frac{\left|e^{-i\lambda_S\Delta t}\left(\frac{1+\epsilon_L}{\sqrt{2}}\right)
\eta_{3\pi^0}
-e^{-i\lambda_L\Delta t}
\left(\frac{1+\epsilon_S}{\sqrt{2}}\right)\right|^2}
{\left|e^{-i\lambda_S\Delta t}\left(\frac{1+\epsilon_S}{\sqrt{2}}\right)
{\eta_{\pi\pi}}
-e^{-i\lambda_L\Delta t}
\left(\frac{1+\epsilon_L}{\sqrt{2}}\right)\right|^2}
\times D_{\CPT} ~
\nonumber\\
&\simeq& |1+2\delta|^2 \left| 1-
\left(
\eta_{3\pi^0}
-{\eta_{\pi\pi}} 
%
\right)
e^{-i(\lambda_S-\lambda_L)\Delta t} \right|^2
\times D_{\CPT}~,
\nonumber\\
&=& |1+2\delta|^2 \left| 1- 
\left(2\delta 
+\epsilon^{\prime}_{3\pi^0} -\epsilon^{\prime}_{\pi\pi}  \right)
e^{-i(\lambda_S-\lambda_L)\Delta t} \right|^2
\times D_{\CPT}~.
\label{ratio4prime}
\end{eqnarray}
\end{linenomath*}
{
It is important to realise from eqs.(\ref{ratio2prime}) and (\ref{ratio4prime}) that there exists a sum rule 
for $\Delta t \gtrsim 0$
given by:
}
\begin{linenomath*}
\begin{eqnarray}
R_{2,\CPT}^{\rm{exp}} ( \Delta t ) +
R_{4,\CPT}^{\rm{exp}} (\Delta t ) &=& 2 D_{\CPT}~,
\end{eqnarray}
\end{linenomath*}
indicating that the quantity $D_{\CPT}$ is measurable within the same experiment.
\par
\begin{figure}[htbp] 
   \centering
   \includegraphics[width=5.4in]{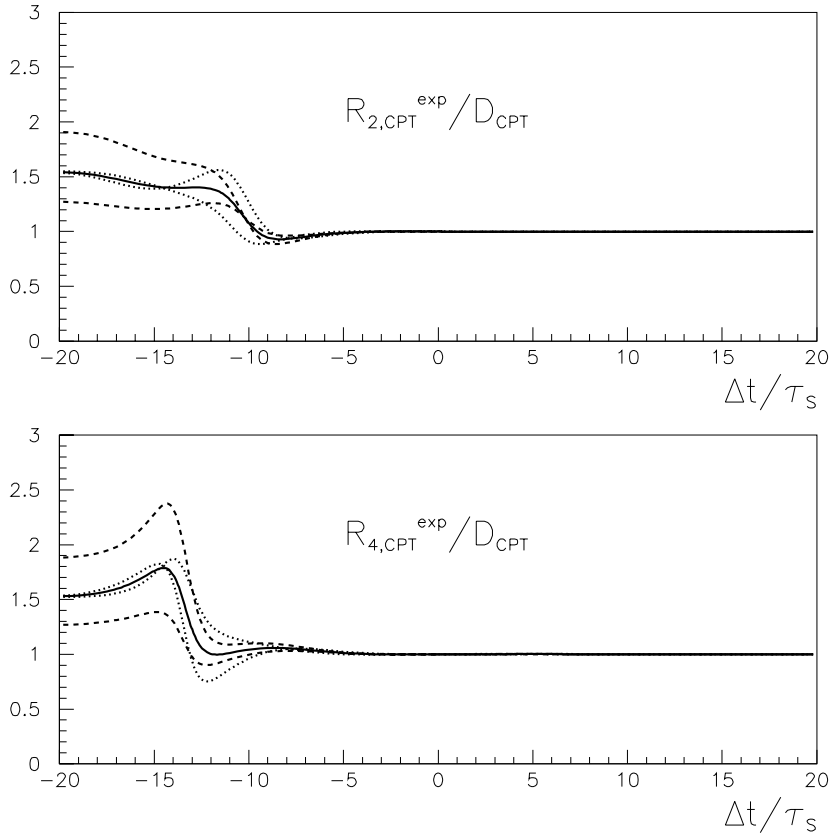} 
   \caption{The expected ratios for the CPT test  $R_{2,\CPT}^{\rm{exp}}(\Delta t)$ (top) and $R_{4,\CPT}^{\rm{exp}}(\Delta t)$ (bottom) 
   as a function of $\Delta t$ (solid line); dashed lines correspond to $\pm10\%$ variation in the absolute value of $\eta_{3\pi^0}$, while dotted lines correspond to a $\pm 10^{\circ}$
variation of its phase (with respect to
the value $\eta_{3\pi^0}=\epsilon_S$).
For visualization purposes 
 the \CPT violating parameters have been fixed to the values $\Re \delta=3.3 \times 10^{-4}$ and $\Im \delta = 1.6 \times 10^{-5}$~.
}
   \label{fig3}
\end{figure}
\begin{figure}[htbp] 
   \centering
   \includegraphics[width=5.5in]{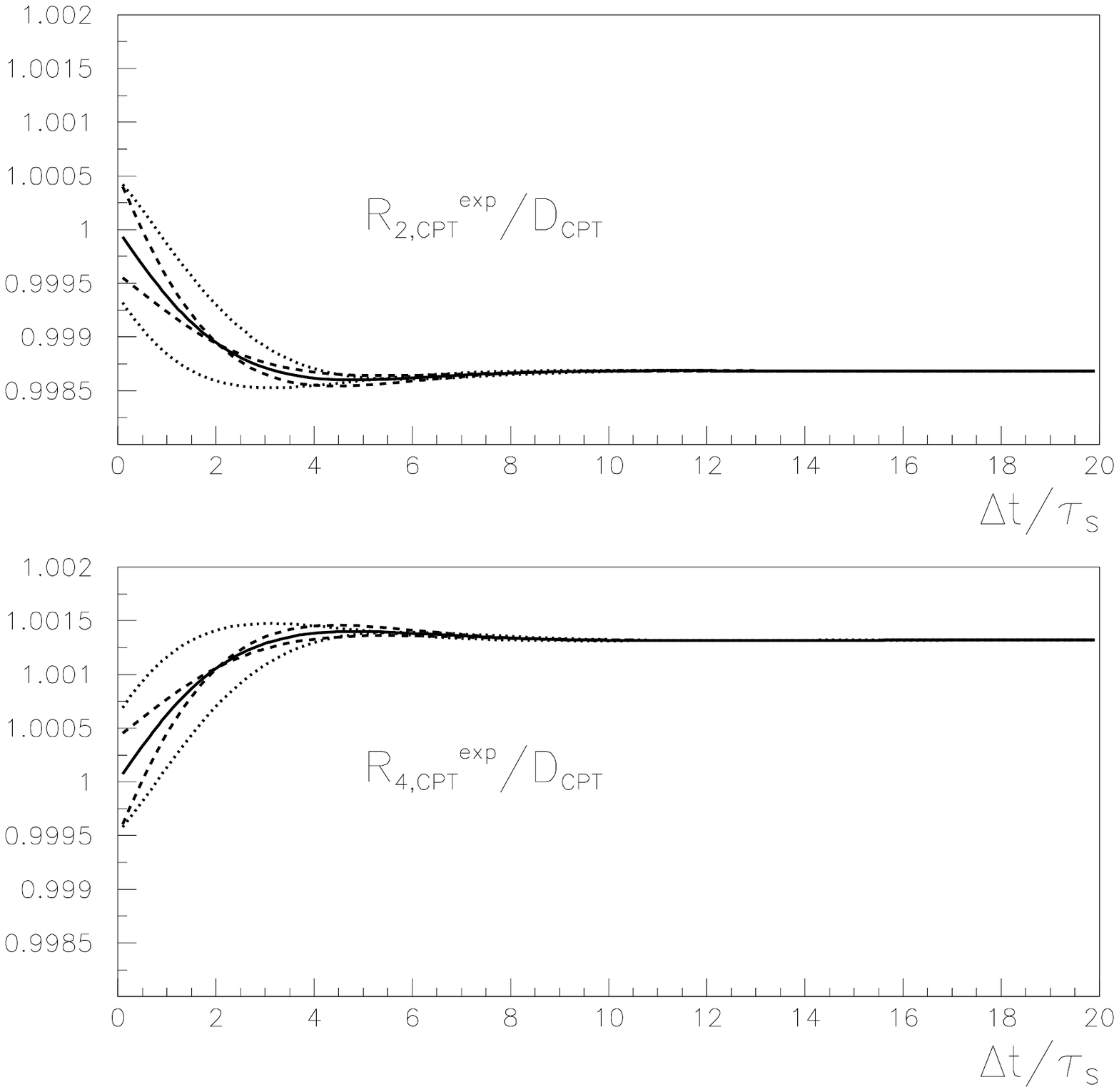} 
   \caption{
   A zoom of the plots shown in fig.\ref{fig3} in the region $0\leq \Delta t \leq20 \tau_S$.
%
}
   \label{fig4}
\end{figure}

{
For $\Delta t =0$ the deviation of 
each ratio 
from unity (once $D_{\CPT}$ is factored out) is only given by the contaminating parameters $\Re \epsilon^{\prime}_{3\pi^0}$ and 
$\Re \epsilon^{\prime}_{\pi\pi}$:
\begin{linenomath*}
\begin{eqnarray}
\label{eq:erreratzero1}
R_{2,\CPT}^{\rm{exp}} ( 0 ) &=& \left[ 1+2\Re (\epsilon^{\prime}_{3\pi^0} -\epsilon^{\prime}_{\pi\pi}  ) \right] \times D_{\CPT}\\
R_{4,\CPT}^{\rm{exp}} (0) &=& \left[ 1-2\Re (\epsilon^{\prime}_{3\pi^0} -\epsilon^{\prime}_{\pi\pi}  ) \right] \times D_{\CPT}~.
\label{eq:erreratzero2}
\end{eqnarray}
\end{linenomath*}
\par
In the limit $\Delta t \gg \tau_S$ there is no dependence on $\eta_{3\pi^0}$ or $\eta_{\pi\pi}$, the limits
 (\ref{eq:tendtoconst1}) and (\ref{eq:tendtoconst2}) are fully recovered:}
\begin{linenomath*}
\begin{eqnarray}
R_{2,\CPT}^{\rm{exp}} ( \Delta t \gg \tau_S) &=& ( 1-4\Re\delta ) \times D_{\CPT}
\label{eq:plateau1}
\\
R_{4,\CPT}^{\rm{exp}} ( \Delta t \gg \tau_S) &=& (1+4\Re\delta ) \times D_{\CPT}~,
\label{eq:plateau2}
\end{eqnarray}
\end{linenomath*}
and any deviation from unity would be a very clean and unambiguous signal of \CPT violation, once $D_{\CPT}$ is factored out.
{
\par
The presence of direct \CP violation contributions 
can mimic \CPT violation effects
mostly in the $\Delta t <0$ region,
while  for $\Delta t \gg \tau_S$ it does not affect the observable ratios (see eqs.(\ref{eq:plateau1}) and (\ref{eq:plateau2})~),
as it can be seen in the
plots of Figures \ref{fig3} and \ref{fig4}. In these plots
the fake effects have been amplified with unrealistic values
of $\eta^{\prime}_{3\pi^0}$ in order to visualise them\footnote{
At lowest order in Chiral Perturbation Theory one has
 ~\cite{ref:wolf3pi,ref:maiani3pi,ref:isidori_cpt} that $\epsilon^{\prime}_{3\pi^0} = -2 \epsilon^{\prime}$, 
 with $\epsilon^{\prime}$ the direct \CP violation parameter in $\pi\pi$ decays.
Therefore
the direct \CP violation contribution is expected to be negligible.
The experimental knowledge on the $\eta_{3\pi^0}$
parameter is much less precise than for
$\eta_{\pi\pi}$, resulting at present in an upper limit:  $\left|\eta_{3\pi^0}\right|< 8.8 \times 10^{-3}$ at 90\% C.L. \cite{ref:KLOE3pi0}.
In the plots of Figs.\ref{fig3} and \ref{fig4} we  considered the contribution of $\epsilon^{\prime}_{3\pi^0}$ enhanced in magnitude
by a large safety factor (of order one hundred)
with respect to the Chiral Perturbation Theory prediction.
},
varying of $\pm10\%$ the absolute value of $\eta_{3\pi^0}$, or of $\pm 10^{\circ}$
its phase, with respect to the
value $\eta_{3\pi^0}=\epsilon_S$.
\par
From these plots one also realises that the time dependences induced by $\delta$ and $\epsilon^{\prime}_{3\pi^0}$
(or $\epsilon^{\prime}_{\pi\pi}$), mostly in the $\Delta t <0$ region,
are different and therefore these parameters could be disentangled, at least in principle.
One should also consider that possible contributions from direct \CPT violation in the decay amplitudes 
in the $\epsilon^{\prime}_{3\pi^0}$ and $\epsilon^{\prime}_{\pi\pi}$ parameters,
cannot be considered fake effects, while instead they are
genuine \CPT violating effects, e.g. responsible of making 
the observable ratios (\ref{eq:erreratzero1}) and (\ref{eq:erreratzero2}) different from unity at $\Delta t =0$.
%
%
}
%
\par
As a conclusion
 the effect of the contamination due to direct \CP violation can be considered fully controllable, and totally irrelevant
 in the limits (\ref{eq:plateau1}) and (\ref{eq:plateau2}).
%
%

\par
If also the $\Delta S = \Delta Q$ rule is not assumed, in analogy with the formalism for the $|\kpp\rangle$
and $|\knn\rangle$ states introduced in Section 2, one should introduce the tagged states which cannot decay into $\ell^-$ or $\ell^+$, respectively, as:
\begin{linenomath*}
\begin{eqnarray}
|\knt\rangle &\equiv & {\rm\widetilde N_0} \left[| \kln\rangle 
- \eta_{\ell^-} |\ksn \rangle \right] 
\\
|\knbt\rangle &\equiv & {\rm\widetilde N_{\bar{0}} } \left[| \kln\rangle 
- \eta_{\ell^+} |\ksn \rangle \right] 
\end{eqnarray}
\end{linenomath*}
%
with
${\rm\widetilde N_0}$,  
$\rm\widetilde N_{\bar{0}} $ two suitable normalization factors,
\begin{linenomath*}
\begin{eqnarray}
\eta_{\ell^+}&=&\frac{ {\langle \ell^+ |T |\kln\rangle} }{ {\langle \ell^+ |T |\ksn\rangle}}
=1 -2 \delta -2 x_+ - 2 x_- \\
\eta_{\ell^-}&=&\frac{ {\langle \ell^- |T |\kln\rangle} }{ {\langle \ell^- |T |\ksn\rangle}}
=-1 -2 \delta +2 x_+^{\star} - 2 x_-^{\star}~,
\end{eqnarray}
\end{linenomath*}
and 
the $x_+$, $x_-$ parameters 
defined as:
\begin{linenomath*}
\begin{eqnarray}
x_{\pm}=\frac{1}{2} \left[ \left(\frac{A(\knb\to \ell^+)}{A(\kn\to \ell^+)} \right) \pm  \left(\frac{A(\kn\to \ell^-)}{A(\knb\to \ell^-)} 
\right)^{\star}\right]~,
\end{eqnarray}
\end{linenomath*}
and corresponding to \CPT-invariant and \CPT-violating $\Delta S = \Delta Q$ rule violation, respectively.
%
\par
The orthogonal partners of $|\knt \rangle$ and $|\knbt \rangle$ states are, respectively:
\begin{linenomath*}
\begin{eqnarray}
|\knbtt\rangle &\equiv & {\rm N_{\bar{0}} } \left[| \kln\rangle 
+\gamma_- |\ksn \rangle \right] 
\\
|\kntt\rangle &\equiv & {\rm N_0} \left[| \kln\rangle 
+\gamma_+ |\ksn \rangle \right] 
\end{eqnarray}
\end{linenomath*}
with
\begin{linenomath*}
\begin{eqnarray}
\gamma_{\pm}=\frac{1-\eta_{\ell_{\pm}}^{\star}\langle\ksn | \kln\rangle}{\eta_{\ell_{\pm}}^{\star} -\langle\kln | \ksn\rangle}~,
\end{eqnarray}
\end{linenomath*}
and
${\rm N_0}$,   
$\rm N_{\bar{0}} $ two suitable normalization factors.
\\
The orthogonal pairs $\{ \knt, \knbtt\}$ and $\{ \knbt, \kntt \}$
(note the different symbols adopted in this case with respect to \kn, \knb)
constitute now the true orthogonal bases to be considered.
%
%
%
The effect of the $\Delta S \neq \Delta Q$ parameters  $x_+$ and $x_-$
can be easily singled out in the explicit expressions of the observable ratios (still neglecting higher order terms in small parameters
and for not too large negative $\Delta t$):
\begin{linenomath*}
\begin{eqnarray}
R_{2,\CPT}^{\rm{exp}} ( \Delta t ) &=&
\frac{P[\knt(0)\to\knn(\Delta t)]}{P[\knnp(0)\to\knbtt(\Delta t)]} \times D_{\CPT} 
\nonumber\\
&=& \frac{\left|e^{-i\lambda_S\Delta t}
(\eta_{\ell^-})
(\eta_{3\pi^0})
-e^{-i\lambda_L\Delta t}
\right|^2}
{\left|e^{-i\lambda_S\Delta t}
({\eta_{\pi\pi}})
-e^{-i\lambda_L\Delta t}
({\eta_{\ell^-}})
\right|^2} 
\times D_{\CPT}~,
\nonumber\\
&\simeq& \left| \frac{1}{\eta_{\ell^-}} \right|^2 \left| 1+ 
\left(\frac{ {\eta_{\pi\pi}} }{\eta_{\ell^-}}-(\eta_{\ell^-})(\eta_{3\pi^0})
\right)
e^{-i(\lambda_S-\lambda_L)\Delta t} \right|^2
\times D_{\CPT}~,
\nonumber\\
&\simeq& |1-2\delta+2 x_+^{\star} - 2 x_-^{\star} |^2 \left| 1+ 
\left(
\eta_{3\pi^0}
- {\eta_{\pi\pi}} 
\right)
e^{-i(\lambda_S-\lambda_L)\Delta t} \right|^2
\times D_{\CPT}~,
\nonumber\\
&=& |1-2\delta+2 x_+^{\star} - 2 x_-^{\star} |^2 \left| 1+ 
\left(
2\delta 
+\epsilon^{\prime}_{3\pi^0} -\epsilon^{\prime}_{\pi\pi}
\right)
e^{-i(\lambda_S-\lambda_L)\Delta t} \right|^2
\times D_{\CPT}~,
\nonumber\\
\label{ratio2ds}
\end{eqnarray}
\end{linenomath*}
\begin{linenomath*}
\begin{eqnarray}
R_{4,\CPT}^{\rm{exp}} ( \Delta t ) &=&
\frac{P[\knbt(0)\to\knn(\Delta t)]}{P[\knnp(0)\to\kntt(\Delta t)]} \times D_{\CPT} 
\nonumber\\
&=& \frac{\left|e^{-i\lambda_S\Delta t}
(\eta_{\ell^+})
(\eta_{3\pi^0})
-e^{-i\lambda_L\Delta t}
\right|^2}
{\left|e^{-i\lambda_S\Delta t}
({\eta_{\pi\pi}})
-e^{-i\lambda_L\Delta t}
({\eta_{\ell^+}})
\right|^2} 
\times D_{\CPT}~,
\nonumber\\
&\simeq& \left| \frac{1}{\eta_{\ell^+}} \right|^2 \left| 1+ 
\left(\frac{ {\eta_{\pi\pi}} }{\eta_{\ell^+}}-(\eta_{\ell^+})(\eta_{3\pi^0})
\right)
e^{-i(\lambda_S-\lambda_L)\Delta t} \right|^2
\times D_{\CPT}~,
\nonumber\\
&\simeq& |1+2\delta+2 x_+ + 2 x_- |^2 \left| 1-
\left(
\eta_{3\pi^0}
- {\eta_{\pi\pi}} 
\right)
%
%
%
e^{-i(\lambda_S-\lambda_L)\Delta t} \right|^2
\times D_{\CPT}
\nonumber\\
&=& |1+2\delta+2 x_+ + 2 x_- |^2 \left| 1-
\left(
2\delta 
+\epsilon^{\prime}_{3\pi^0} -\epsilon^{\prime}_{\pi\pi}
\right)
%
%
%
e^{-i(\lambda_S-\lambda_L)\Delta t} \right|^2
\times D_{\CPT}~,
\nonumber\\
\label{ratio4ds}
\end{eqnarray}
\end{linenomath*}
\par
{
In the limit $\Delta t =0$ the deviation of each ratio  from unity (once $D_{\CPT}$ is factored out) is given by the contaminating parameters $\Re \epsilon^{\prime}_{3\pi^0}$, 
$\Re \epsilon^{\prime}_{\pi\pi}$, and $x_+$, and by the $x_-$ parameter, which is explicitly \CPT violating 
in the $\Delta S \neq \Delta Q$ decay amplitudes, and can be considered a genuine source of \CPT violation:
}
\begin{linenomath*}
\begin{eqnarray}
R_{2,\CPT}^{\rm{exp}} ( 0 ) &=& \left[ 1+2\Re (\epsilon^{\prime}_{3\pi^0} -\epsilon^{\prime}_{\pi\pi}  ) 
+4\Re (x_+ - x_-  )
\right] \times D_{\CPT}\\
R_{4,\CPT}^{\rm{exp}} (0) &=& \left[ 1-2\Re (\epsilon^{\prime}_{3\pi^0} -\epsilon^{\prime}_{\pi\pi}  ) 
+4\Re (x_+ + x_-  )
\right] \times D_{\CPT}~.
\end{eqnarray}
\end{linenomath*}
In the limit $\Delta t \gg \tau_S$ we get:
\begin{linenomath*}
\begin{eqnarray}
R_{2,\CPT}^{\rm{exp}} ( \Delta t \gg \tau_S) &=& ( 1-4\Re\delta +4\Re x_+ - 4 \Re x_- ) \times D_{\CPT}\\
R_{4,\CPT}^{\rm{exp}} ( \Delta t \gg \tau_S) &=& (1+4\Re\delta +4\Re x_+ + 4 \Re x_- ) \times D_{\CPT}~.
\end{eqnarray}
\end{linenomath*}
\par
These results suggest the possibility of having 
a measurement independent of $x_+$ and $D_{\CPT}$,
directly measuring the double ratio:
\begin{linenomath*}
\begin{eqnarray}
\frac{ R_{2,\CPT}^{\rm{exp}} ( \Delta t) }{
R_{4,\CPT}^{\rm{exp}} ( \Delta t) } 
&\simeq& 
\left( 1-8\Re \delta -8 \Re x_- \right) 
\left| 1+ 2
\left(
\eta_{3\pi^0}
- {\eta_{\pi\pi}} 
\right)
e^{-i(\lambda_S-\lambda_L)\Delta t} \right|^2
\nonumber\\
&=&
\left( 1-8\Re \delta -8 \Re x_- \right) 
\left| 1+ 2
\left(
2\delta 
+\epsilon^{\prime}_{3\pi^0} -\epsilon^{\prime}_{\pi\pi}
\right)
e^{-i(\lambda_S-\lambda_L)\Delta t} \right|^2~,
\nonumber\\
\label{doubleratiods}
\end{eqnarray}
\end{linenomath*}

which becomes for $\Delta t =0$:
\begin{linenomath*}
\begin{eqnarray}
\frac{ R_{2,\CPT}^{\rm{exp}} ( 0) }{
R_{4,\CPT}^{\rm{exp}} (0) }= 1 -8 \Re x_- +4\Re (\epsilon^{\prime}_{3\pi^0} -\epsilon^{\prime}_{\pi\pi}  ) ~.
 \label{doubleratiodszero}
\end{eqnarray}
\end{linenomath*}

and in the limit $\Delta t \gg \tau_S$:
\begin{linenomath*}
\begin{eqnarray}
\frac{ R_{2,\CPT}^{\rm{exp}} ( \Delta t \gg \tau_S) }{
R_{4,\CPT}^{\rm{exp}} ( \Delta t \gg \tau_S) }= 1-8\Re\delta -8 \Re x_- ~.
 \label{doubleratiodslimit}
\end{eqnarray}
\end{linenomath*}

{
The double ratio (\ref{doubleratiods}) constitutes one of the most robust observables for our proposed \CPT test.
In the limit $\Delta t \gg \tau_S$ it exhibits a pure and genuine \CPT violating effect, even without the assumptions 
of the validity of the $\Delta S = \Delta Q$  rule and of negligible contaminations from direct \CP violation.
\\
In principle, the possible contribution of $\Re x_-$ might
be disentangled from the one of $\Re\delta$, or at least bound, by making a study of the time dependence 
of the double ratio in the $\Delta t<0$ region -- the most difficult experimentally, due to the lack of statistics -- and 
independently measuring the direct \CP violation contributions in $3\pi^0$ and $\pi\pi$ decays,
or making an ansatz on their size.
}
%
\par
{
The KLOE-2 experiment at the DA$\Phi$NE facility could make a measurement of the two observable ratios
$R_{2,\CPT}^{\rm{exp}} ( \Delta t)$ and $R_{4,\CPT}^{\rm{exp}} ( \Delta t)$,
with an integrated luminosity $L$ of
$\mathcal{O}(10 \hbox{ fb}^{-1}  )$~\cite{kloe2epjc}.
The $I(f_1,f_2 ; \Delta t)$ distributions 
have been evaluated with a simple Monte Carlo simulation,
making the approximation of a gaussian $\Delta t$ experimental resolution
with $\sigma=1~\tau_S$, and a full detection efficiency,
as discussed in detail in Ref.\cite{tviol}. From this study it emerges that 
the $I(\ell^{\pm}, 3\pi^0; \Delta t)$ distributions, at the considered  integrated luminosity,
have very few or no events  for $\Delta t \lesssim-5~\tau_S$.
While a complete feasibility study is beyond the scope of the present paper, 
by considering a large 
$\Delta t$ interval in the statistically most populated region,
e.g. 
$0\leq \Delta t \leq 300~\tau_S$,
which is the most interesting one for our \CPT test,
a statistical sensitivity on the double ratio (\ref{doubleratiodslimit})
of $(3.0\times 10^{-3})$, $(2.1\times 10^{-3})$, and $(1.5\times 10^{-3}$)
is obtained 
for $L=$~5, 10, and 20~$\hbox{ fb}^{-1} $, respectively.
Once
translated into an uncertainty  
on $\Re\delta$, these results might 
improve
the present
best measurement by CPLEAR \cite{cplearred}.

}

\par
As a final remark it's worth noting that the double ratio in eq.(\ref{doubleratiodslimit}) in practice corresponds to the following ratio of combinations of the semileptonic asymmetries for \ksn and \kln~:
\begin{linenomath*}
\begin{eqnarray}
\frac{ R_{2,\CPT}^{\rm{exp}} ( \Delta t \gg \tau_S) }{
R_{4,\CPT}^{\rm{exp}} ( \Delta t \gg \tau_S) } =\frac{1+A_L}{1-A_L}\times\frac{1-A_S}{1+A_S} 
\simeq 1+2(A_L-A_S)~,
\label{cpttest4}
\end{eqnarray}
\end{linenomath*}
with 
\begin{linenomath*}
\begin{eqnarray}
A_S&=&\frac{\Gamma(\ksn\to \ell^+)-\Gamma(\ksn\to \ell^-)}{\Gamma(\ksn\to \ell^+)+\Gamma(\ksn\to \ell^-)}
\nonumber\\
A_L&=&\frac{\Gamma(\kln\to \ell^+)-\Gamma(\kln\to \ell^-)}{\Gamma(\kln\to \ell^+)+\Gamma(\kln\to \ell^-)}~.
\label{eq:asyms}
\end{eqnarray}
\end{linenomath*}
Eq.(\ref{cpttest4}) is model independent in the interpretation of the \CPT-violating asymmetries.
Therefore the same observable could be accessible by 
measuring separately and independently $A_S$ and $A_L$, instead of the double ratio at large $\Delta t$ (\ref{doubleratiodslimit}). This is indeed a part of the KLOE-2 program~\cite{kloe2epjc}.

\section{Conclusions}
\label{conclusions}
\par
 A novel \CPT test has been studied in the neutral kaon system based on
the direct comparison of a transition probability with 
its 
\CPT reverse transition. The appropriate preparation and detection of 
{\it in} and {\it out} states in both the reference and the reverse processes is 
made by exploiting the EPR entanglement of neutral kaons produced in a
$\phi$-factory and using their decays as filtering measurements of the kaon
states only.
\par
   We have built the time dependence of two independent ratios of
observable quantities, 
which relate
 to a given transition and to its
\CPT-transformed, 
by selecting the two semileptonic decays for flavour tag,
the $\pi\pi$ and $3\pi^0$ decays for \CP tag, and the two time orderings
of the decay pairs. These observables have been interpreted in the
standard Weisskopf-Wigner approach by means of the \CPT violating $\delta$
parameter in the mass matrix of $\kn-\knb$~. The dynamical \CPT-asymmetry is
generated with the time evolution, the observable ratios being 
unity at $\Delta t=0$, and for $\Delta t \gg \tau_S$ beyond the interference
region, where one has the best statistics of events, the deviation from
unity is given by $\Re \delta$, a genuine \CPT violating parameter independent of
$\Delta \Gamma$, for which the decay is not an essential ingredient.
\par
    We have demonstrated that the necessary knowledge of the decay
properties is controllable and some strategies, like a sum rule and the
ratio of the observable ratios, can be used. Furthermore possible
spurious effects induced by \CP violation in the decay and/or a violation
of the $\Delta S= \Delta Q$ rule have been shown to be either negligible or
disentangled by the dependence with the time evolution. The proposed
measurement is thus fully robust leading to definite conclusions on
\CPT Violation.
\\
These observables would be able to be measured in the KLOE-2
experiment at the DA$\Phi$NE facility in Frascati, 
with 
a statistical precision of $\mathcal{O}(10^{-3})$.
%
%

\section{Acknowledgements}
This research has been supported by MINECO and Generalitat Valenciana
 Projects FPA 2011-23596 and GVPROMETEO II 2013-017 and by Severo Ochoa
 Excellence Centre Project SEV 2014-0398.



\end{document}